\documentstyle[PASJadd]{PASJ95}
\draft
\begin{document}

\title{Superwind-Driven Intense H$_2$ Emission in NGC 6240}

\author{Youichi {\sc Ohyama},$^1$ Michitoshi {\sc Yoshida},$^2$
Tadafumi {\sc Takata},$^3$ Masatoshi {\sc Imanishi},$^{1, 4}$\\
Tomonori {\sc Usuda},$^3$ Yoshihiko {\sc Saito},$^5$
Hiroko {\sc Taguchi},$^6$ Noboru {\sc Ebizuka},$^7$\\
Fumihide {\sc Iwamuro},$^8$ Kentaro {\sc Motohara},$^8$
Tomoyuki {\sc Taguchi},$^8$ Ryuji {\sc Hata},$^8$\\
Toshinori {\sc Maihara},$^9$ Masanori {\sc Iye},$^1$
Toshiyuki {\sc Sasaki},$^3$ George {\sc Kosugi},$^3$\\
Ryusuke {\sc Ogasawara},$^3$ Junichi {\sc Noumaru},$^3$
Yoshihiko {\sc Mizumoto},$^1$ Masafumi {\sc Yagi},$^1$\\
and Yoshihiro {\sc Chikada}$^1$
\\[12pt]
\small$^1$ {\it National Astronomical Observatory of Japan, 2-21-1 Osawa,
Mitaka, Tokyo 181-8588}\\
{\it E-mail(YO): ohyama@optik.mtk.nao.ac.jp}\\
\small$^2$ {\it Okayama Astrophysical Observatory, NAOJ, Kamogata-cho,
Okayama 719-0232}\\
\small$^3$ {\it Subaru Telescope, NAOJ, 650 A'ohoku Place, Hilo, HI 96720, USA}\\
\small$^4$ {\it Institute for Astronomy, University of Hawaii, Honolulu, HI
96822, USA}\\
\small$^5$ {\it Department of Astronomy, University of Tokyo, Bunkyo-ku,
Tokyo 113-0033}\\
\small$^6$ {\it Department of Astronomy and Earth Sciences, Tokyo Gakugei
University, Koganei, Tokyo 184-8501}\\
\small$^7$ {\it Communications Research Laboratory, Koganei, Tokyo
184-8795}\\
\small$^8$ {\it Department of Physics, Kyoto University, Sakyo-ku, Kyoto
606-8502}\\
\small$^9$ {\it Department of Astronomy, Kyoto University, Sakyo-ku, Kyoto
606-8502}}

\abst{We have performed a long-slit K band spectroscopic observation of the
luminous infrared galaxy NGC 6240.
Spatially extended H$_2$ emission is detected over 3.3 kpc around the
two nuclei.
The peak position of the H$_2~v=1-0~S$(1) emission in the slit is
located $\sim 0.\hspace{-2pt}''3 - 0.\hspace{-2pt}''4$ north of the southern
nucleus.
It is almost the midpoint between the southern nucleus and the peak
position of the $^{12}$CO ($J=1-0$) emission.
Based on the line-ratio analyses, we suggest the excitation mechanism of
H$_2$ is pure thermal at most positions.
In the northern region including the northern nucleus, the H$_2$
velocity field shows only a slight velocity gradient and is not highly
disturbed.
On the other hand, the kinematics of the H$_2$ emitting gas is more
complicated around the southern nucleus and its south region.
In the southern region we find the following three velocity components
in the H$_2$ emission:
the blueshifted shell component ($\approx -250$ km s$^{-1}$ with
respect to $V_{\rm sys}$) which is recognized as a distinct C-shape
distortion in the velocity field around the southern nucleus, the
high-velocity blueshifted ``wing'' component ($\sim -1000$ km s$^{-1}$
with respect to $V_{\rm sys}$), and the component indicating possible
line splitting of $\sim 500$ km s$^{-1}$.
The latter two components are extended to the south from the southern
nucleus.
We show that the kinematic properties of these three components can be
reproduced by expanding motion of a shell-like structure around the
southern nucleus.
The offset peak position of the H$_2$ emission can be understood if we
assume that the shell expanding to the north interacts with the
extragalactic molecular gas which has been transfered during the
course of the merging of the two nuclei.
At the interface between the shell and the molecular gas concentration
the cloud-crushing mechanism proposed by Cowie et al. (1981)
may work efficiently, and the intense H$_2$ emission is thus expected
there.
With this mechanism, the H$_2$ luminosity can be explained without
global shock driven by the collision of the two nuclei.
All these findings lead us to propose a model that the most H$_2$
emission is attributed to the shock excitation driven by the superwind
activity of the southern nucleus.}

\kword{Infrared: galaxies --- Infrared: spectra --- Galaxies: individual
(NGC 6240) --- Galaxies: interacting --- Galaxies: intergalactic medium
--- Shock waves}

\maketitle
\thispagestyle{headings}

% section 1
\section{Introduction}

NGC 6240 (= IRAS 16504+0228 = UGC 10592 = IC 4625) is one of the
famous luminous infrared galaxies (LIGs:
$L$(IR)$=L$($8-1000\mu$m)$=4.6\times 10^{11}\LO$: Sanders
\& Mirabel 1996) at a distance of 98 Mpc (Heckman et al.\ 1987).
It is a merging system containing two radio and near-infrared nuclei
(Condon et al.\ 1982; Fried \& Schulz 1983; Eales et al.\ 1990; Thronson
et al.\ 1990; Herbst et al.\ 1990; Colbert, Wilson, \& Bland-Hawthorn
1994; Sugai et al.\ 1997; Tacconi et al.\ 1999).
Although this galaxy is often called as a prototypical LIG (Wright,
Joseph, \& Meikle 1984; Joseph \& Wright 1984; Heckman, Armus, \&
Miley 1987, 1990; Armus, Heckman, \& Miley 1990; Klaas et al.\ 1997),
it is also well known as an object with unusually luminous
near-infrared molecular hydrogen (H$_2$) emission (e.g., Rieke et
al. 1985; DePoy, Becklin, \& Wynn-Williams 1986).
Several H$_2$ excitation mechanisms have been proposed to date
including 1) thermal excitation driven by shocks (Rieke et al.\ 1985;
DePoy et al.\ 1986; Lester, Harvey, \& Carr 1988; Elston \& Maloney
1990; Herbst et al.\ 1990; van der Werf et al.\ 1993; Sugai et
al. 1997), 2) UV fluorescence (Tanaka, Hasegawa, \& Gatley 1991), 3) soft
X-ray heating (Draine \& Woods 1990; see also Mouri et al.\ 1989), and
4) formation pumping (Mouri \& Taniguchi 1995).
Recently, Sugai et al. (1997) presented their new high-quality
spectrum and clearly showed that the H$_2$ spectrum can be interpreted
as a pure thermal excitation.
Although the origin of the H$_2$ emission in LIGs and ultraluminous
infrared galaxies (ULIGs: Sanders \& Mirabel 1996) is generally
attributed to their intense star-formation activities at their nuclei
and/or active galactic nuclei (AGNs) (e.g., Moorwood \& Oliva 1990;
Goldader et al.\ 1995), NGC 6240 is often considered to be an
exceptional case.
It is often discussed that the global shock caused by a galaxy-galaxy
collision brings its huge H$_2$ luminosity.
This is because
1) current star-formation activity (or, current supernova explosion
rate) is not intense enough to reproduce its huge H$_2$ luminosity
(e.g., Rieke et al.\ 1985; Draine \& Woods 1993),
2) unusually large intensity ratio of H$_2~v=1-0~S$(1) [hereafter,
$1-0~S$(1)] to Br$\gamma$ ($\simeq 45$: van der Werf et al.\ 1990)
cannot be explained by normal star-formation activities (e.g., Rieke
et al.\ 1985; Draine \& Woods 1993),
and 3) the peak position of the H$_2$ emission is located not at each
nucleus but between the two nuclei where no star-forming activity is
detected (Herbst et al.\ 1990; van der Werf et al.\ 1993; Sugai et
al. 1997).
However, it is important to remember that these kinds of evidence are
not direct ones, i.e., we do not have any spatial and kinematic
information directly suggesting that the intense H$_2$ emission does
come from the galaxy-galaxy collision interface.

Although thermal excitation driven by the galaxy-galaxy collision
might be a main excitation mechanism of the strong H$_2$ emission,
there remains some possibilities of other excitation mechanisms.
Like other normal LIGs, it seems very likely that intense starburst
activity and/or AGN also contribute more or less to the H$_2$
emission.
In fact, there are many pieces of evidence suggesting intense
starburst activity in this galaxy (Wright et al.\ 1984; Rieke et
al. 1985; Smith, Aitken \& Roche 1989; van der Werf et al.\ 1993).
Also there are many pieces of evidence suggesting an activity of
``superwind'', which is a galaxy scale blastwave driven by numerous
supernova explosions (Tomisaka \& Ikeuchi 1989; Heckman et al.\ 1990;
Suchkov et al.\ 1994), and some previous
authors have indeed suggested its importance to the H$_2$ emission
(Elston \& Maloney 1990; Herbst et al.\ 1990; van der Werf et
al. 1993).
X-ray heating from an AGN, which is recently found by the hard X-ray
observation (Iwasawa \& Comastri 1998), might also contribute to the
H$_2$ emission (Mouri et al.\ 1989).
In this way, the physical mechanism of the H$_2$ emission seems to be
very complicated in this galaxy.

In order to study the origin of the H$_2$ emission in more detail and
to understand the nature of the activity of NGC 6240, it is important to
obtain a spatial information of the emission-line properties.
Thus, we have conducted a long-slit K-band spectroscopic observations
of this galaxy with Subaru 8.2m telescope (Kaifu 1998) in order
to perform detailed spectroscopic analyses.

% section 2
\section{Observation and Data Reduction}

Observations were performed on the night of 1999 April 29 during the
test observation phase of Subaru telescope at the top of Mauna Kea,
Hawaii, using the grism spectroscopy mode of CISCO (Cooled Infrared
Spectrograph and Camera for OHS: Motohara et al.\ 1998).
The detector used is a $1024 \times 1024$ Hawaii array with a projected
pixel size of $0.\hspace{-2pt}''115$ along the slit and 8.6\AA~ at 2.3$\mu$m
along the dispersion direction.
Slit width is $0.\hspace{-2pt}''5$ and slit length is about 2 arcminutes, which
is long enough to obtain sky background information in each on-source
frame.
The slit position angle was set to the position angle of the two
nuclei ($19^\circ$: Herbst et al.\ 1990; Thronson et al.\ 1990) in order
to observe the northern and southern nuclei (hereafter, the N nucleus
and the S nucleus, respectively) as well as the intergalactic region
at a same time (Figure 1
\footnote{
This image was obtained with the CISCO imaging mode with a K band
filter through a wide $2''$ slit with the position angle of
$19^\circ$.
Note that because this image was taken just for the source
acquisition, the dark subtraction and the flatfielding were not
applied.
It is presented just for the reader to easily understand the rough
source positions along the slit.
The image is averaged over 2 pixels ($0.\hspace{-2pt}''23$).
}
).
% Fig. 1

The total exposure time was 1500 second for on source (fifteen exposures,
each of 100 second integration).
The object was placed on only two positions along the slit during
exposures.
A nearby A3-type star SAO 122007 and a K0-type star SAO 122106 were also
observed just before observing NGC 6240 in order to correct for
atmospheric absorption.
Seeing size around K band wavelength were $0.\hspace{-2pt}''5-0.\hspace{-2pt}''6$
during the observations.
Dome flat spectra were obtained at another observing run on 1999 June
25 with the same instrumental setup.

Data reduction was performed with IRAF
\footnote{
IRAF is distributed by the National Optical Astronomy Observatories,
which are operated by the Association of Universities for Research in
Astronomy, Inc., under cooperative agreement with the National Science
Foundation.
}.
First, the dark is subtracted from each frame.
Each dark-subtracted frame was divided by the dark-subtracted dome
flat spectrum.
Then, wavelength calibration of each frame was made using the telluric
OH emission lines to an accuracy of $\sim 2.6$\AA.
The spectral resolution measured from the OH emission lines was $\sim
26$\AA~ FWHM at 2.2$\mu$m ($350$ km s$^{-1}$).
We have measured the wavelength of the OH lines in the
wavelength-calibrated sky spectra and found that the systematic error
is less than 1\AA~ around 2.2$\mu$m ($14$ km s$^{-1}$).
Background sky emission was removed by interpolating the adjacent sky
spectra with a linear interpolation.
The wavelength calibrated frames were combined using 3 $\sigma$
clipping averaging method after shifting the images along the slit.
The spectra of the standard stars were also reduced in the same way.
Br$\gamma$ absorption feature of the A-type star was removed in the
following way.
First, the spectrum of the A-type star was divided by that of the
K-type star in order to cancel out the common features such as the
atmospheric absorption and the instrumental sensitivity.
The resultant ratio spectrum shows an isolated Br$\gamma$ absorption
feature and this feature was removed using Voigt profile fitting.
Then, the ratio spectrum was multiplied by the spectrum of the K-type
stars, giving modified A-type star spectrum without Br$\gamma$
absorption feature.
The modified spectrum was used for correcting the atmospheric
absorption features and instrumental sensitivity assuming a 8590 K
blackbody spectrum.
Because the sky condition was not photometric during the observations,
no absolute flux calibration was made.

% section 3
\section{Results}

% section 3.1
\subsection{Broadband image}

Broad K band image (Figure 1) clearly shows two nuclei (the N and the
S nuclei) at a separation of $\simeq 1.\hspace{-2pt}''9$ with the position
angle of $\simeq 19^\circ$.
This result is consistent with the previous studies in K band (Eales
et al.\ 1990; Herbst et al.\ 1990; Thronson et al.\ 1990; Sugai et
al. 1997), although Herbst et al. (1990) reported a slightly smaller
separation of the two nuclei ($1.\hspace{-2pt}''52\pm 0.\hspace{-2pt}''02$).

One must be careful when comparing the nuclear positions seen in the
radio continuum and the K band image.
Colbert et al. (1994) and Tacconi et al. (1999) found the double
nuclei in the radio continuum emission.
Although the position angles of the two radio nuclei
($19^{\circ}\hspace{-4.5pt}.\hspace{.5pt}7$ and
$21^\circ$ at 8.3 GHz and 5 GHz, respectively) are consistent with one
of the K band nuclei ($19^\circ$), the distance between the two is
significantly smaller in radio ($\simeq 1.\hspace{-2pt}''53$ and
$1.\hspace{-2pt}''7$ at 8.3 GHz and 5 GHz, respectively) than in K band
($\sim 1.\hspace{-2pt}''9$).
This apparent discrepancy can be understood as an effect of the dust
extinction.
The visual extinction ($A_{\rm v}$) of this galaxy inferred from the
large column density ($N_{\rm H_2} \gtsim 2 \times 10^{23}$
cm$^{-2}$) is as large as $\gtsim 200$ mag (Tacconi et al.\ 1999),
indicating heavy dust extinction even at K band ($A_{\rm k} \simeq
0.112 \times A_{\rm v}\gtsim 22$ mag; Binney \& Merrifield 1998).
Thus the true nuclei should be at the extinction-free radio nuclei,
i.e., the true N and S nuclei are located at slightly south and north
of the apparent K band positions ($\sim 0.\hspace{-2pt}''1-0.\hspace{-2pt}''2$),
respectively.

% section 3.2
\subsection{The spectra and the flux distribution}

Spatially extended H$_2$ emission is detected over 7$''$ region
(3.3 kpc) around the two nuclei.
Figure 2 shows the K band spectra of NGC 6240 at various positions.
Rich H$_2$ line features including $1-0~S$(0), $1-0~S$(1),
$1-0~S$(2), $1-0~S$(3), $2-1~S$(1), and $2-1~S$(3) emission lines are
detected at most positions.
% Fig. 2
The $2-1~S$(2) and possibly $2-1~S$(4) lines are also detected at some
positions.
Weak Br$\gamma$ emission is detected at both the N and the S nuclei.
Deep stellar CO absorption is also detected at both nuclei, being
consistent with the report by Lester \& Gaffney (1994).

The peak position of $1-0~S$(1) is located at about $0.\hspace{-2pt}''
3-0.\hspace{-2pt}''4$ north-north-east of the S nucleus (Figure 3) and is much
closer to the S nucleus than to the N nucleus.
% Fig. 3
Recent high resolution $^{12}$CO($J=2-1$) mapping shows that the CO
peak position is located at $0.\hspace{-2pt}''6$ north-north-east of the S
nucleus seen in the radio continuum images (Tacconi et al.\ 1999) and
is marked in Figure 1.
Thus, the peak position of $1-0~S$(1) is located at almost the
midpoint between the CO peak position and the S nucleus.
This result is inconsistent with the previous works based on
narrow-band emission-line imagings,
showing that intense $1-0~S$(1) emission arises from the region
between two nuclei with its peak around the midpoint of the two
(Herbst et al.\ 1990; van der Werf et al.\ 1993; Sugai et al.\ 1997).
Note, however, that the overall spatial extent of the H$_2$ emission found
in our data is nearly consistent with the previous results.
The equivalent width of the $1-0~S$(1) emission is the lowest ($20-40$\AA) at
both nuclei, where the continuum emission is the strongest, and is larger
at other regions ($\gtsim 100$\AA).

% section 3.3
\subsection{The velocity field}

The heliocentric velocities and the line widths of the $1-0~S$(1)
emission measured with a single Gaussian fitting at various positions
are shown in Figure 3.
A continuum-subtracted peak-normalized $1-0~S$(1) spectroscopic image
is also shown in Figure 4.
% Fig. 4
The velocity is nearly constant and is approximately $7300$ km
s$^{-1}$ in the northern region including the N nucleus and its north.
We find the systemic velocity $V_{\rm sys}$ of NGC 6240 $\simeq 7300$
km s$^{-1}$ from compilation of the previously published infrared and
radio observations (the previous measurements of $V_{\rm sys}$ are
summarized in Table 1.).
% Table.1
Thus, the radial velocity of the H$_2$ emitting gas coincides with the
systemic one and the gas motion is relatively quiescent in the
northern region.

On the other hand, we find a remarkable velocity variation in the southern
region:
the emission line is significantly blueshifted and the velocity field
is highly disturbed to form a C-shaped distortion over $\sim
1''$ (450 pc) region around the S nucleus (see Figure 4).
At $0.\hspace{-2pt}''3 - 0.\hspace{-2pt}''5$ south from the S nucleus the
velocity reaches down to $-250$ km s$^{-1}$
($V_\solar\simeq 7050$ km s$^{-1}$; see Figure 3) with respect to
$V_{\rm sys}$.
It must be remembered that the true position of the S nucleus after
correcting for the extinction between the two nuclei is expected to be
located about $\sim 0.\hspace{-2pt}''1 - 0.\hspace{-2pt}''2$ north of the

apparent K band nucleus.
Thus, the center of the C-shaped distortion almost corresponds
spatially to the true S nuclear position.
It is noteworthy that about 50\% of the $1-0~S$(1) emission covered in
our slit comes from this blueshifted region.
This kind of a peculiar velocity field has been previously found by
Elston \& Maloney (1990), but they could not resolve the detailed
structure of the velocity field because of the limited spatial and
velocity resolution of their spectrum.
We first revealed the detailed dynamical structure of the
circumnuclear H$_2$ emission-line region of NGC 6240.

This remarkable C-shaped velocity field would account for the
difference between the flux peak position of $1-0~S$(1) measured by
this work and that derived from the previous narrow band imagings.
As pointed out in the previous section, we find that the peak position
of $1-0~S$(1) is located much closer to the S nucleus than previously
reported positions.
How can we understand this difference ?
Using narrow band imagings, Sugai et al. (1997) showed that the peak
of $1-0~S$(1) is located between the two nuclei.
Their images were taken with three Fabry-Perot settings at $V_{\rm
sys}$ ($7339$ km s$^{-1}$ in their paper), $V_{\rm sys}-175$ km
s$^{-1}$, and $V_{\rm sys}+175$ km s$^{-1}$.
Comparing with our results, it seems likely that the most blueshifted
$1-0~S$(1) emission ($\sim 7000-7100$ km s$^{-1}$) could not be
detected in their images.
Based on their three narrow-band images (centered at $V_\solar=6910, 7390$,
and $7470$ km s$^{-1}$), van der Werf et al. (1993)
showed that the position of the $1-0~S$(1) peak moves from south to
north with increasing velocity, although the peak position of the
velocity-integrated $1-0~S$(1) emission comes around the midpoint of
the two nuclei.
The blueshifted $1-0~S$(1) emission found in our data ($V_{\rm
sys}-250$ km s$^{-1}$) should be detected in their blue-band image in
which the peak is closer to the S nucleus, being consistent with our
result.
Herbst et al. (1990) found a southwest extension of the $1-0~S$(1)
emission ($\ltsim 0.\hspace{-2pt}''5$ from the S nucleus) in addition to the
component around the midpoint of the two nuclei
\footnote{
Although Sugai et al. (1997) emphasized that the peak position of the
$1-0~S$(1) emission is located between the two nuclei, we point out
that their contour plot also shows an extension toward the southwest
direction (see their Figure 1).
This extension seems to correspond to the southwest extension found in
Herbst et al. (1990).
}.
Since our slit position angle is $19^{\circ}$, it seems likely that
our slit covers a part of this extension.
If this component is blueshifted with respect to $V_{\rm sys}$, then
it would be observed as a blueshifted component around the S nucleus
in our data.
In these ways, we find that our result is consistent with the previous
results.

In addition to the blueshifted emission around the S nucleus, we find
a high-velocity blueshifted emission-line ``wing'' whose velocity
exceeds $-1000$ km s$^{-1}$ with respect to the peak velocity of the
profile (Figure 4).
This kind of a profile was previously found by van der Werf et
al. (1993).
With our spatially resolved spectra, we can investigate the spatial
distribution of this high-velocity wing component and find that such a
component extends around the S nucleus and ti its southern regions
($\sim 2''$, or $\sim 950$ pc).

The line width is nearly constant at the region between the two nuclei
and around the N nucleus ($550-600$ km s$^{-1}$ FWHM after correcting
for the instrumental line broadening), which is consistent with the
previous results (e.g., van der Werf et al.\ 1993).
It is well known that this galaxy shows unusually broader line width
($\simeq 550$ km s$^{-1}$) compared with other galaxies with the
intense H$_2$ emission (e.g., van der Werf et al.\ 1993).
We newly find that the line width at $1'' -2''$ south of the
S nucleus is even broader ($700-800$ km s$^{-1}$ FWHM) than more
northern regions
\footnote{
A reason why previous studies reported relatively narrow line width
may be that the line width tends to be narrower at stronger H$_2$
emission region and that they could only measure the
intensity-weighted line width because of their insufficient spatial
resolution.
}.
We also find that the line profile at this region shows boxy profile
(with nearly flat-topped profile and relatively weak low intensity
wing) rather than an usual Gaussian-like profile seen in more northern
regions (Figure 4).
In order to understand this peculiar line profile, we try to represent
the observed spectra with the combination of two velocity components
each of which is blue- and redshifted with respect to the mean central
velocity measured with the single Gaussian fitting (Figure 3).
We assume Gaussian line profile with the line width of $550$ km
s$^{-1}$ FWHM for each component.
For simplicity, the intensities of the two components are set to be
equal because of the nearly symmetric line profile (except for the low
intensity blueshifted high-velocity wing component).
The central velocity is set to $V_{\rm sys}- 100$ km s$^{-1}$ from
Figure 3.
The fitting was made by eye.
As a result, we find that superposition of the two components with a
velocity difference of $500$ km s$^{-1}$ can represent the observed
line profile at $7''$ south of the S nucleus (Figure 5).
% Fig. 5
We thus propose a model that two broad ($550$ km s$^{-1}$ FWHM)
emission lines whose velocity difference is $\approx 500$ km s$^{-1}$
make the boxy profile seen in the southern region.
Note that the velocity difference of the two components ($500$ km
s$^{-1}$) is nearly twice the velocity of the C-shaped velocity
distortion observed around the S nucleus ($250$ km s$^{-1}$).

The line width of the Br$\gamma$ emission is nearly the same as that
of the $1-0~S$(1) emission (i.e., $\sim 500-600$ km s$^{-1}$ FWHM) at
all positions with a detectable Br$\gamma$ emission.
We find no kinematic evidence for the presence of a broad line region
of AGN ($\gtsim 5000$ km s$^{-1}$ FWHM; Osterbrock 1989).

% section 3.4
\subsection{Emission-line ratios and the excitation mechanism}

We examine the H$_2$ excitation mechanism at various positions.
Following Mouri (1994), we show a line-ratio diagnostic diagram
for discussing the excitation mechanisms in Figure 6.
% Fig. 6
Some data points scatter around a theoretical locus of the thermal
excitation with a temperature of about 2000 K.
Others show slightly lower $1-0~S$(2)/$1-0~S$(0) ratio and scatter
around observed data points of supernova remnants (SNRs).
It is known that temperature gradient within the shock front causes
the ratio slightly lower than this locus in SNRs (e.g., Beckwith et
al. 1983).
Thus, the molecular gas is thermally excited at most positions.
On the other hand, we find no evidence for the non-thermal excitation
with the larger $2-1~S$(1)/$1-0~S$(1) ratio ($\gtsim 0.5$) anywhere.

We also show population diagrams at various positions (Figure 7).
% Fig. 7
We find that the data points at each position are aligned along a
single straight line on this diagram, indicating that thermal
excitation is dominated at most positions.
The region at $1'' - 1.\hspace{-2pt}''2$ south of the S nucleus marginally
shows relatively stronger $1-0~S$(1) and weaker $2-1~S$(3) emissions.
Figure 8 shows a spatial variation of the $v=2-1$ vibration
temperature ($T_{\rm vib}$) and $v=1$ rotation temperature ($T_{\rm
rot}$).
% Fig. 8
Most regions show a temperature of $\sim 2000$ K, being consistent
with the results of Sugai et al. (1997).
In order to assess a possible contribution of other excitation
mechanisms, we show a plot of the ratio of $T_{\rm vib}$ to $T_{\rm
rot}$ as a function of slit positions (Figure 8).
The ratio should be unity in case of a pure thermal excitation and be
slightly larger in shocks with a temperature gradient (which is
empirically $\simeq 1.25 = 1/0.8$; Tanaka et al.\ 1991).
In case of an UV fluorescent, the ratio becomes much larger than unity
since $T_{\rm vib}$ ($\sim 6000$ K) is much larger than $T_{\rm rot}$
($\sim 1000$ K) (e.g., Tanaka et al.\ 1991).
Although the ratio is larger than unity at some positions, it can be
understood as a case with a temperature gradient at most positions.
The $1-0~S$(1) emission at $1'' - 1.\hspace{-2pt}''2$ south of the S
nucleus shows slightly enhanced ratio of $T_{\rm vib}$ to $T_{\rm
rot}$ although it is still consistent within errors with the thermal
excitation with a temperature gradient.
However, because the depression of the $2-1~S$(3) emission is also
observed as well as the enhanced $1-0~S$(1) emission, UV fluorescent
might contribute the H$_2$ emission to some extent there.

% section 4
\section{Discussion}

% section 4.1
\subsection{Superwind origin of the H$_2$ emission}

We find the three velocity components in the H$_2$ emission in the
southern region of NGC 6240:
1) the blueshifted component ($\approx -250$ km s$^{-1}$ with respect
to $V_{\rm sys}$) which is recognized as a distinct C-shape distortion
of the velocity field,
2) the high-velocity ``wing'' component ($\sim -1000$ km s$^{-1}$ with
respect to $V_{\rm sys}$), and 3) the component indicating possible
line splitting of $\sim$ 500 km s$^{-1}$.
In the following sections we focus on these kinematic properties and
discuss the origin of the extremely intense H$_2$ emission of NGC 6240
in the course of a ``superwind'' hypothesis.

% section 4.2
\subsubsection{Evidence for superwind}

First we discuss the origin of the high-velocity ``wing'' component
($-1000$ km s$^{-1}$ with respect to $V_{\rm sys}$).
It is difficult for the galaxy-galaxy collision at a speed of $75-200$
km s$^{-1}$ (van der Werf et al.\ 1993) to produce a high-velocity
material at a speed up to $\sim 1000$ km s$^{-1}$ (van der Werf et
al. 1993).
Moreover, such a component is only found around the S nucleus and the
southern region, not around the region between the two nuclei where
the two galaxies are likely to collide with each other.
Thus, the galaxy-galaxy collision would not be responsible for this
component.
On the other hand, superwind origin of such a high-velocity material
is promising since broad optical emission lines are detected around
nuclear 1 kpc region in this galaxy and is attributed to the superwind
on the basis of morphological, kinematic, and spectral evidence
(Heckman et al.\ 1990).

The possible line splitting seen in the southern region can also be
attributed to the superwind activity.
Such type of the velocity field is often observed in the region of the
expanding shell-like structure or biconical outflow seen in most
superwind galaxies (e.g., Heckman et al.\ 1990).
Most previous authors noted the extended south-south-west component
$1''-2''$ away from the S nucleus (Elston \& Maloney 1990;
Herbst et al.\ 1990; van der Werf et al.\ 1993).
Because this region is located at another side of the interface
between the two colliding galaxies and no energy source is detected
such as a star-formation activity there, superwind activity would be a
most plausible agent for exciting the molecular gas in this region.
The region showing line splitting corresponds spatially to this
component, suggesting that a superwind indeed affects the molecular
gas at this region.
Slight blueshift in this region ($-100$ km s$^{-1}$) can be understood
if we assume that the expanding direction of the bubble is not within
the sky plane but is slightly tilted to our line-of-sight.
It is interesting that the CO emission also extends toward this
direction and shows broader emission line profile like $1-0~S$(1)
emission (Tacconi et al.\ 1999), indicating that both of the cold
(traced by CO) and the warm (traced by $1-0~S$(1)) molecular gas is
affected by the superwind activity.

The C-shaped velocity field around the S nucleus strongly support the
idea that the H$_2$ emitting gas is affected by the superwind
activity.
There are some nearby examples showing cold gas (atomic and molecular
gas) outflow at a speed of several hundred km s$^{-1}$ (e.g., Phillips
1993; Nakai et al.\ 1987; Ohyama \& Taniguchi 1997).
Theoretical calculations also predict such an expanding structure
(e.g., Suchkov et al.\ 1994).
It is expected that some of the gas entrained in the high-velocity
ionized gas outflow would be heated up by the shock to form a shell
with the intense H$_2$ emission.
If there is an expanding shell around the S nucleus, the redshifted
component would be obscured by the heavy dust extinction around the S
nucleus.
As noted before, the dust extinction is severe even in K band ($A_{\rm
k} \gtsim 22$ mag), which is dusty enough to obscure the H$_2$
emission from behind the nucleus and only blueshifted velocity
structure is expected to be observed
\footnote{
Although the peak position of the molecular gas is located around
$0.\hspace{-2pt}''6$ north of the S nucleus, significant fraction of the gas is
distributed around the S nucleus (Tacconi et al.\ 1999).
We expect that the column density around the S nucleus is still enough
to obscure the redshifted part of the H$_2$ emission.
}
, i.e., the velocity field would be C-shaped.
This type of a velocity structure is one of the common characteristics
seen in galaxies with superwinds (e.g., Heckman et al.\ 1990).

Since it is well known that this galaxy actually exhibits a superwind
activity (e.g., Heckman et al.\ 1987, 1990; Armus et al.\ 1990; Keel
1990), it is natural to attribute these three velocity components to
the entrained and shocked molecular gas within the superwind.
Because both of the high-velocity wing component and the line
splitting is not detected around the N nucleus, the starburst and the
superwind of the S nucleus would be responsible for exciting these
components.
This interpretation is supported by the fact that the
[Fe\,{\footnotesize II}]$\lambda$ 1.64$\mu$m emission, which is considered
to arise at a shock driven by supernova explosions
(van der Werf et al.\ 1993; Sugai et al.\ 1997), is much stronger at the S
nucleus than at the N nucleus.
All these considerations lead us to conclude that most H$_2$ emission
in the southern region comes from the H$_2$ gas entrained in and
shocked by the superwind blowing from the S nucleus.

Recently, Tacconi et al. (1999) conducted a high-resolution
$^{12}$CO($J=2-1$) mapping.
They found a velocity gradient along the position angle of $\sim
45^\circ$ and showed the presence of blueshifted ($\sim -350$ km
s$^{-1}$ with respect to their $V_{\rm sys}=7320$ km s$^{-1}$)
molecular gas around the S nucleus.
The trend of this gradient is similar to that found in the
$1-0~S$(1) emission (van der Werf et al.\ 1993).
The peak position of the blueshifted CO emission ($V_\solar=
6805-7005$ km s$^{-1}$) is located at just $\sim 0.\hspace{-2pt}''2$ north of
the S nucleus, indicating the presence of the blueshifted cold
molecular gas around the S nucleus.
Thus, it is very likely that some part of the cold molecular gas
traced with the CO emission is heated up by the shock driven by the
superwind to emit blueshifted $1-0~S$(1) emission around the S
nucleus.

%section 4.3
\subsubsection{Evidence against galaxy-galaxy collision}

In spite of many pieces of evidence for the superwind origin of the
H$_2$ emission, we find no evidence for the galaxy-galaxy collision
origin of the strong $1-0~S$(1) emission.
In the collision model, kinetic energy released by the global
collision of the two nuclei is converted to $1-0~S$(1) emission.
Thus, most intense $1-0~S$(1) emission is expected to arise from the
collision interface.
We, however, find that the peak of the emission is located much closer
to the S nucleus, being inconsistent with the model prediction.
In addition to the discussion on the flux distribution, the kinematic
information gives another clue against the collision model.
We find a C-shaped velocity distortion of the H$_2$ emission at a
maximum blueshifted velocity of $-250$ km s$^{-1}$ around the S
nucleus.
On the other hand, the stellar velocity difference between the two
nuclei is less than $75$ km s$^{-1}$ (Lester \& Gaffney 1994),
indicating that only emission line component is blueshifted around the
S nucleus
\footnote{
The stellar heliocentric velocity reported by Lester \& Gaffney (1994)
is $V_\solar=7275 \pm 50$ km s$^{-1}$ and is consistent with the H$_2$
emission velocity measured around the N nucleus.
We thus confirm that the emission line component around the S nucleus is
blueshifted with respect to the stellar one.
}.
Thus, any models in which the H$_2$ emission is associated with the
stellar component of the S nucleus can be rejected.
Moreover, although some kind of a violent dynamical structure is
expected at the interface of the two nuclei in the collision model,
such as a sudden velocity change and/or a large velocity dispersion,
such evidence is found neither around the region between the two nuclei
where two galaxies are likely to be colliding nor other regions along the slit.
In spite of these difficulties in the collision model, a superwind
model can easily explain the velocity field, i.e., an expanding
shell-like structure whose redshifted part is obscured by the heavy
dust extinction on the extended H$_2$ gas with a nearly flat velocity
field.
Hence, we argue that the galaxy-galaxy collision would not be a
main agent of the strong $1-0~S$(1) emission of NGC 6240.
Note, however, that we cannot reject a possibility of the galaxy-galaxy
collision model since all pieces of evidence described above are not
direct ones.

% section 4.4
\subsubsection{Spatial and dynamical structure of the superwind}

Theoretical calculations predict that a superbubble is formed at the
early stage of the superwind evolution and that it will break out to
form a biconical structure when the bubble extends large enough
compared with the scaleheight of the surrounding material (Tomisaka \&
Ikeuchi 1988; Heckman et al.\ 1990; Suchkov et al.\ 1994).
The superbubble is originally spherical around the starburst nucleus
and will elongate into the direction where the density gradient is the
largest (Tomisaka \& Ikeuchi 1988; Suchkov et al.\ 1994).
This direction is usually perpendicular to the dense disk gas in case
of the normal nuclear starburst in spiral galaxies.
In the case of NGC 6240, however, this would not be the case since the
surrounding matter is expected to be highly disturbed in the course of
the merging process, leading to the complicated superwind structure as
observed (e.g., Heckman et al.\ 1987; Armus et al.\ 1990; Keel 1990).
Thus, it is important to discuss the superwind morphology and the
velocity structure in detail in order to know the distribution of the
ambient matter and to understand how the superwind interacts with this
gas.

As pointed out before, the emission line structure is not symmetric
around the S nucleus.
Rather, the emission is extended farther to the south of the S nucleus
($\sim 3''$, or 1.4 kpc) than to the north ($\sim 1''$, or
450 pc).
At the south region, we find a kinematic evidence for the expanding
structure (i.e., the high-velocity wing component and the line
splitting).
It is difficult to say clearly whether the southern region shows a
closed bubble-like structure or an open conical structure because of
the low surface intensity of the $1-0~S$(1) emission.
At the north of the S nucleus, the emission line velocity goes back to
$V_{\rm sys}$ at just $\sim 1''$ north of the S nucleus.
A compact expanding bubble can explain such a property in which the
bubble is expanding to the north in the sky plane and the
line-of-sight expanding velocity becomes to be zero at the top of the
bubble.
Hence, a highly asymmetric elongated bubble/wind expanding to
north-south direction within the sky plane would be a most likely
picture of the superwind of NGC 6240.

Here we compare the observed size and velocity of the bubble with that
of the theoretical expectations.
Following Heckman et al. (1996), we assume an idealized model of a
single spherical bubble within the uniform surrounding medium ($n_0$).
The kinetic energy ($L_{\rm mech}$) is assumed to be injected at a
constant rate during a time ($t$) and the radiative losses are
negligible.
In this case, the radius ($r$) and the expanding velocity ($v$) can be
expressed as
% equation 1
\begin{equation}
L_{\rm mech}=8\times 10^{42} L_{\rm bol, 11}~{\rm ergs~s}^{-1},
\end{equation}
% equation 2
\begin{equation}
r=7 L_{\rm mech, 43}^{1/5}n_{0, -2}^{-1/5}t_7^{3/5} ~{\rm kpc},
\end{equation}
% equation 3
\begin{equation}
v=410 L_{\rm mech, 43}^{1/5}n_{0, -2}^{-1/5}t_7^{-2/5} ~{\rm km~s}^{-1},
\end{equation}
where $L_{\rm mech, 43}$ is the kinetic energy injection rate in units
of $10^{43}$ ergs s$^{-1}$, $L_{\rm bol}, 11$ is the bolometric
luminosity in units of $10^{11}$ $\LO$, $n_{\rm 0, -2} $ is
the number density of the ambient gas in units of $10^{-2}$ cm$^{-3}$,
and $t_7$ is the time in units of $10^7$ years (Heckman et al.\ 1993,
1996).
The time ($t$) is estimated as $r_{\rm bubble}/v_{\rm expansion}
\simeq 500$ pc $/ 250$ km s$^{-1} \simeq 2\times 10^6$ years,
indicating relatively young age of the bubble (see for other examples
of the young superwind; Yoshida, Taniguchi, \& Murayama 1999 and
references therein).
The bolometric luminosity is taken from Wright et al. (1984) and is
$L_{\rm bol}=5.0\times 10^{11} \LO$ (converted to our adopted
distance).
Substituting these values into the equations, we estimate that $r =
3.5, 2.2, 1.4, 0.88, 0.56$ kpc and $v = 1030, 650, 410, 260, 160$ km
s$^{-1}$ for $n_0=10^{-2}, 10^{-1}, 10^0, 10^1, 10^2$ cm $^{-3}$,
respectively.
Comparing with the observation, we find that the observation can be
well represented (within a factor of two) by the model with
$n_0=10^{1-2}$ cm$^{-3}$.
Although the inferred density is higher than that of the usual
extragalactic ambient matter ($\sim 10^{-2}$ cm$^{-3}$), it can be a
natural consequence of the smaller size of the bubble ($r \simeq 500$
pc) compared with the total size of the entire galaxy ($r \simeq 30$
kpc: Heckman et al.\ 1987).
The asymmetric structure of the bubble can be understood as a result
of the density gradient of the ambient matter, i.e., smaller radius
toward north direction where the ambient gas density is higher and
vice versa.
This result is consistent with the fact that the CO is concentrated
toward north direction of the S nucleus.
In this way, the superbubble model can reproduce the observed
properties very well.
It is interesting to point out that the age of the bubble ($\sim
2\times 10^6$ years) is much younger than that of Arp 220 ($\sim
3\times 10^7$ years: Heckman et al.\ 1996).

Previous narrow-band H$_2$ imaging studies have revealed an extended
complex H$_2$ emission-line structure around the two nuclei
(Herbst et al.\ 1990; van der Werf et al.\ 1993; Sugai et al.\ 1997) and
it is very difficult to imagine a simple bubble-like structure in
them.
Optical narrow-band imaging studies also show extended ($\sim 60\times
50$ kpc: Heckman et al.\ 1987) filamentary emission-line nebula (see
also Armus et al.\ 1990; Keel 1990).
Inhomogeneity of the surrounding material would be a main reason for
these complex structures.
However, we point out that two prominent tail-like features of the
H$_2$ emission are extending toward southeast and southwest directions
from the S nucleus.
These structures are most clearly visible in an image of Sugai et
al. (1997; see their Figure 1).
If we assume that the tails imply the projected edges of a cone, the superwind
seems to be blowing toward south with an opening angle of $\sim
90^\circ$.
Further discussions on the detailed structure of the superwind would
require deeper emission-line images.

It seems interesting to point out that the size of the bubble we find
($\sim 1$ kpc) is much smaller than that of the large-scale optical
nebula ($\sim 60\times 50$ kpc) although both are likely to be driven
by numerous supernova explosions in the nuclear region.
In order to produce the large-scale nebula within the timescale
for a small bubble ($\sim 2\times 10^6$ years), unusually large
expanding velocity ($> 10000$ km s$^{-1}$) is required.
Thus, the superwind in this galaxy seems to comprise at least two
bubbles/winds with different ages, i.e., the young compact bubble and
the surrounding older larger bubble/wind, although the number of
bubbles is not certain.
The age of the larger bubble/wind is estimated to be $8\times 10^7$
years under an assumption of the common expanding velocity for both the
bubbles/winds ($250$ km s$^{-1}$).
In order to make such a multiple bubble system, starburst events must
have occurred twice, i.e., $\sim 2\times 10^6$ years and $\sim 8\times
10^7$ years ago.
We expect that the starburst activity has been triggered by the tidal
dynamical disturbance several times when the separation of the two
nuclei becomes the smallest while they are orbiting around each other.

Theoretical models predict that an expanding bubble should brake out
because of Rayleigh-Taylor instability when a bubble expands into an
outer low density halo (Tomisaka \& Ikeuchi 1988; Heckman et al.\ 1990;
Suchkov et al.\ 1994).
If the density of the ambient matter is high enough or the energy
supply from the nuclear region is small enough, instability is
suppressed and the ionized gas would be confined not to form a
biconical structure.
Since starburst activity in NGC 6240 is relatively high
[$L$(IR)$=4.6\times 10^{11}$ $\LO$], it is expected that the
bubble would be broken out unless the density of the ambient matter is
unusually high.
However, NGC 6240 might be an unusual case with an extraordinary high
density extragalactic matter.
It is well known that the huge amount of the molecular gas is
distributed between the two nuclei, rather than at each nucleus, as a
result of the dynamical disturbance caused by the galaxy-galaxy merger
(Bryant \& Scoville 1999; Tacconi et al.\ 1999).
If the molecular gas accumulates between the nuclei before the
superbubble begins to expand ($10^{7-8}$ years after the onset of the
starburst), the expanding superbubble toward the north of the S
nucleus would be blocked by this dense extranuclear molecular gas and
the broken-out of the bubble would be inhibited, forming a compact
bubble-like structure as observed.
On the other hand, the superbubble would expand more freely to the
south direction.
If this is the case, we can explain the reason why the peak position
of the $1-0~S$(1) emission comes closer to the S nucleus than the peak
position of the CO emission since the shock front between the bubble
and the ambient matter should come between the two.

In this model, the H$_2$ emission in the northern region is difficult
to be explained.
Judging from a complex morphology and velocity field seen also in this
region (van der Werf et al.\ 1993), we would argue that the most plausible
agent is a superwind.
Starburst and/or AGN activities of the N nucleus, rather than that of
the S nucleus, might be responsible for the superwind activity.
In fact, the activity of the N nucleus is indicated by
[Fe\,{\footnotesize II}]$\lambda 1.64\mu$m emission associated with it
(van der Werf et al.\ 1993).
Since the superwind activity seems to have occurred at least twice, it
seems also possible that the northern H$_2$ emission is excited by the
shock driven by the older superwind, rather than by the current one which is
responsible for the C-shaped velocity distortion around the S nucleus.
We cannot discuss this possibility further only with our data.

% section 4.5
\subsubsection{Can superwind explain the huge H$_2$ luminosity?}

Most previous authors discussed that the galaxy-galaxy collision is
the main energy source of the intense H$_2$ emission in NGC 6240
because the starburst activity would not be strong enough to reproduce
its huge H$_2$ luminosity [$L$($1-0~S$(1))$ \simeq 1 \times 10^8$
$\LO$; Rieke et al.\ 1985; Draine \& Woods 1990; van der Werf
et al.\ 1993].
However, one must remember that the previous estimates of the
$1-0~S$(1) luminosity assume an ordinary nuclear starburst.
We emphasize that such estimates would not be applicable to the case
of NGC 6240 in which the superwind interacts with the dense
intergalactic medium.

If we adopt a canonical value of the conversion factor of kinetic
energy supplied from supernova explosions to H$_2$ emission ($\sim
0.02$), required supernova rate is as high as 15 supernovae per year
(Draine \& Woods 1990).
The supernova rate in NGC 6240 is estimated to be only $3.2-4.8$ if we
simply scale up the supernova rate of the nearby prototypical
starburst galaxy M82 ($\sim 0.2-0.3$ SNRs per year; Rieke et
al. 1980; Kronberg, Biermann, \& Schwab 1985) with their far-infrared
luminosities [$L$(FIR)$=3.3\times 10^10$ $\LO$ for M82 and
$5.2\times 10^{11}$ $\LO$ for NGC 6240: Heckman et al.\ 1990]
\footnote{
Based on the extensive modeling of the starburst properties of M82 and
NGC 6240, Rieke et al. (1985) showed that the starburst models for the
two galaxies are similar, suggesting that the supernova rate scales with
the far-infrared luminosity between the two galaxies.
}.
As pointed out by Draine \& Woods (1993), however, the efficiency can
be as large as 0.2 if the medium is clumpy and the cloud-crushing is
efficient.
If the cloud-intercloud density constant is $\simeq 10^2$, the
cloud-crushing will produce shock waves in the clouds with a shock
speed of $30-50$ km s$^{-1}$ where $1-0~S$(1) emission is efficiently
produced when the blastwave speed is $300-500$ km s$^{-1}$ (Cowie,
McKee, and Ostriker 1981; see also Elston \& Maloney 1990).
As noted before, NGC 6240 seems to be a rare case with a huge amount
of the intergalactic matter.
It is easy to imagine that this matter is disturbed and its spatial
distribution becomes highly clumpy in the course of the merging
process.
It is important to remember that this galaxy shows unusually broad
line width ($\sim 550$ km s$^{-1}$ FWHM), suggesting that the
cloud-crushing mechanism might be very efficient.
We thus argue that the $1-0~S$(1) luminosity can be explained only by the
superwind activity.

Another key point in discussing the origin of the huge H$_2$
luminosity is its unusually large $1-0~S$(1)/Br$\gamma$ intensity
ratio.
The ratio is the largest one observed so far in LIGs (Goldader et
al. 1995).
Although the ratio has been reported to be as large as $\sim 45$ (van der
Werf et al.\ 1990), we find the ratio to be only $\simeq 10$ around the S
nucleus.
The main reason for this discrepancy would be the data quality.
The high signal-to-noise ratio, high spatial resolution spectra of
ours enable us to measure the weak Br$\gamma$ feature more accurately
than the previous works (Remember that Br$\gamma$ is detected only
around the S and the N nuclei while peak of the H$_2$ emission is
offseted from the nuclei).
Although this ratio ($\simeq 10$) is still very large, it can now be
explained by normal star-formation models.
Mouri \& Taniguchi (1992) calculated the $1-0~S$(1)/Br$\gamma$ ratio
for various starburst models and found that the ratio ranges $0.1-10$.
The ratio becomes larger in models with the lower upper mass limit of
the initial mass function.
This is because kinetic energy exciting H$_2$ released from supernova
explosions does not change with the mass of their progenitors, while
ionizing photons from O-type stars decrease rapidly with the decrease
of the mass of the most massive stars.
Also, the ratio becomes larger in models at a post-starburst phase
because O-type stars supplying most ionizing photons are gone.
Thus, the O-type star deficient stellar population can explain the
observed ratio.
In fact, some evidence for the O-type star deficient stellar population of
this galaxy was found (Draine \& Woods 1990; Tanaka et al.\ 1991).
Note, however, that our result only indicates that the superwind-driven
shock is enough to explain the H$_2$ luminosity and does not necessarily
rule out the contribution of a global shock caused by the galaxy-galaxy
collision.

Here one might have a question on why only NGC 6240 shows unusually
strong H$_2$ emission.
One of the possible reasons is that the chance probability for the
superwind to interact with the extragalactic dense matter is very small.
NGC 6240 should be a rare case in which the direction of the mass
transfer during the merging process and that of the outflow of the
superwind are nearly the same.
Another possible reason is the short timescale of the H$_2$ emitting
phase during the evolution of this galaxy.
If the expansion velocity of the superbubble is $\sim 250$ km s$^{-1}$
as suggested by the maximum line-of-sight velocity of the C-shaped
velocity distortion around the S nucleus, the shock front would pass
through the core of the CO emission ($\sim 500$ pc: Tacconi et
al. 1999) in only $\sim 2\times 10^6$ years.
As pointed out before, the age of the bubble is also very young ($\sim
2\times 10^6$ years).
Because these timescales are significantly smaller than that of the
starburst ($\sim 10^{7-8}$ years), the intense H$_2$ emission would
fade out soon and only weaker H$_2$ emission associated with the
nuclear star formation and/or AGN activities would be observed like
other ordinary LIGs after $\gtsim 2\times 10^6$ years.
It seems very likely that the coincidence of both two special conditions
make NGC 6240 be a very rare object with an exceptionally strong H$_2$
emission.

% section 4.6
\subsection{Physical conditions of the H$_2$ gas}

The excitation mechanism of the H$_2$ emission of NGC 6240 has long
been a controversy.
Based on a spectrum presented by Lester et al. (1988), several
excitation mechanisms besides the thermal excitation have been proposed.
Two important characteristics of the Lester et al.'s spectrum are
1) the rotation temperature is lower than the vibration temperature
($T_{\rm rot}=1300$ K and $T_{\rm vib}=2860$ K)
and 2) $2-1~S$(3) emission is unusually weak and only upper limit is
reported.
Tanaka et al. (1991) discussed that the UV fluorescence mechanism is
significant in order to explain the first characteristic.
They pointed out that the reported upper limit of the $2-1~S$(3) flux
is too small and they are not serious on the second characteristic.
On the other hand, Draine \& Woods (1990) emphasized the importance of
the second characteristic and favored the X-ray heating mechanism.
The formation pumping mechanism is also discussed in order to
reproduce the peculiar spectrum (Mouri \& Taniguchi 1995).

Recently, Sugai et al. (1997) presented their high quality H$_2$
spectrum.
Surprisingly, their spectrum significantly differs from that of Lester
et al. (1988) and can be interpreted as a pure thermal excitation
because they found neither an excess of the vibration temperature nor
weaker-than-expected $2-1~S$(3) emission
($T_{\rm rot} \simeq T_{\rm vib}=1910$ K; Sugai et al. 1997).
Why the two spectra are different from each other?
One possible reason is the different aperture positions.
Sugai et al. (1997) placed their slit at the midpoint of the two
nuclei.
At this region, our data indicates a thermal excitation, being
consistent with Sugai's result.
On the other hand, Lester et al. (1988) placed their aperture at ``the
position of peak signal'' which can be interpreted that they placed
their aperture closer to the S nucleus which is brighter than the N
nucleus.
One evidence suggesting the aperture difference is the equivalent
width of the $1-0~S$(1) emission:
Comparing with their spectra, we find that the equivalent width
measured by Lester et al. (1988) is smaller than that of Sugai et
al. (1997).
Since the equivalent width changes dramatically with positions and it
becomes smaller around the S nucleus (Figure 3), it seems very likely
that Lester et al. (1988) indeed placed their aperture closer to the S
nucleus.
In fact, our spectrum at $+0.\hspace{-2pt}''46$ north of the S nucleus, where
the $1-0~S$(1) flux is the maximum with an equivalent width
$\simeq 30-70$\AA, looks similar to the spectrum of Lester et al. (1988)
that shows the $1-0~S$(1) equivalent width $62$\AA.
Our spectrum at $+1.\hspace{-2pt}''15$ north of the S nucleus where equivalent
width is the maximum ($\simeq 100$\AA) looks similar to the spectrum
of Sugai et al. (1977) (see Figures 2 and 3).
Unfortunately, however, because of the smaller equivalent widths of
emission lines and strong stellar absorption features, we cannot
confirm the two characteristics of Lester et al.'s spectrum around the
S nucleus.
Because [Fe\,{\footnotesize II}] emission is detected at the S nucleus and
supernova explosions are the most probable agents (van der Werf et al.\ 1993;
see also Sugai et al.\ 1997), it is expected that the intense starburst
around the S nucleus supplies huge amount of UV photons, leading to
additional UV fluorescent excitation as well as the shock heating
(Tanaka et al.\ 1991).
The UV photons may also be responsible for a possible UV fluorescent
excitation found at $\sim 0.\hspace{-2pt}''5$ south of the S nucleus.

We find no evidence for the X-ray heating (Draine \& Woods 1990) since
$2-1~S$(3) is detected except around the S nucleus.
Also we find no excitational evidence for the AGN activity anywhere.
Thus, we conclude that the main excitation mechanism is the thermal
excitation caused by the shock heating at most positions with a
possible additional contributions of UV fluorescent and/or X-ray
heating around the S nucleus.

% section 5
\section{Proposed Scenario of NGC 6240 System}

Here we summarize a proposed scenario of the merging and superwind
evolution in NGC 6240.

1) Two gas-rich galaxies (the N and the S nuclei) merge.
A huge amount of molecular gas is forced to fall onto the two
nuclei as a result of the dynamical disturbance during the merging
process.
At the same time, the dynamical disturbance triggers the nuclear
starburst at each nucleus.
The starburst activity is stronger at the S nucleus.
A schematic picture at this stage is shown in Figure 9(a).
% Fig. 9
The nuclear starburst activity was triggered several times while the
two nuclei are orbiting around each other.

2) When $10^{7-8}$ years passed since the onset of the starburst,
the massive stars begin to explode and a superbubble starts to expand.
As the superbubble expands, it interacts with the dense and perturbed
molecular gas at the northern region of the S nucleus and heats it
up.
Stronger-than-normal H$_2$ emission lines are observed there as a
result of the efficient cloud-crushing mechanism.
Both the molecular gas that was supplied during the merging process and
the material blown-off by the superwind would emit strong H$_2$
emission lines.
At the same time, the bubble expands more freely toward the south
direction and forms an asymmetrically elongated superbubble.
A schematic picture at this stage is shown in Figure 9(b).
Multiple bubbles are formed as a result of multiple triggerings of the
starburst.

3) The strong H$_2$ emission would be observed only for $2\times 10^6$ years
in which the expanding superbubble interacts with the extragalactic dense
matter.
Then NGC 6240 would be observed as a normal LIGs with a
moderate H$_2$ emission.

% section 6
\section{Summary}

We have performed spatially-resolved spectroscopic analyses of the
H$_2$ emission lines in NGC 6240.
We find that most of the H$_2$ emission is attributed to the shock driven by
the superwind of the S nucleus, rather than by the conventional global
shock driven by the collision of the two nuclei.
Our main conclusions are listed below.

1) The peak position of the H$_2$ $1-0~S$(1) emission is not at
    the midpoint of the two nuclei.
Rather, it is located much closer to the S nucleus.

2) In the southern region (the region around the S nucleus and south of
   the nucleus) we identify the following three velocity components in the
   H$_2$ emission:
a) the blueshifted shell component ($\approx -250$ km s$^{-1}$ with
   respect to $V_{\rm sys}$) which shows a distinct
   C-shape distortion in the velocity field around the southern
   nucleus,
b) the high-velocity blueshifted ``wing'' component ($\sim -1000$ km
   s$^{-1}$ with respect to $V_{\rm sys}$), and
c) the component indicating possible line splitting of $\sim 500$ km
   s$^{-1}$.
The latter two components extend to the south from the S nucleus.

3) The presence of these three velocity components can be explained naturally
   by the superwind activity of the S nucleus.
On the other hand, we found no evidence for the classical galaxy-galaxy
   collision origin of the $1-0~S$(1) emission, although we cannot reject
   the possibility completely.

4) The huge $1-0~S$(1) luminosity can be explained only by the
   superwind-driven shock heating, if we assume a higher conversion
   efficiency of the kinetic energy of SNRs to the $1-0~S$(1) emission
   through the efficient cloud-crushing mechanism.
This mechanism efficiently works at the shock interface between the
   expanding superbubble and the extranuclear molecular gas since the
   molecular gas is likely to be highly clumpy and is dynamically
   disturbed during the merging of the two nuclei.

5) The excitation condition of the H$_2$ emission lines is the shock
   heating at most positions except for around the S nucleus where
   additional UV fluorescent excitation is expected.

6) We propose an evolutional scenario of NGC 6240 in which
   the galaxy-galaxy merger drives the molecular gas to accumulate
   between the two nuclei and triggers a starburst activity at each
   nucleus.
The intense $1-0~S$(1) emission comes from the molecular gas entrained in
   and shocked by the superwind which starts to blow $10^{7-8}$ after
   years since the onset of the starburst.
\par
\vspace{1cm}\par
We would like to thank all the people in the Subaru project.
This research has been done using the facilities at the Astronomical
Data Analysis Center of the National Astronomical Observatory, Japan,
which is an inter-university research institute of astronomy operated
by Ministry of Education, Science, Culture, and Sports.

%------------------------------------------------------------------------------
% References
%------------------------------------------------------------------------------

\clearpage
\section*{References}

\re
Armus, L., Heckman, T.M., Miley, G.K.\ 1990, ApJ 364, 471
\re
Baan, W.A., Haschick, A.D., Buckley, D., Schmelz, J.T.\ 1985, ApJ 293, 394
\re
Beckwith, S., Evans, N.J., II, Gatley, I., Gull, G., Russell, R.W.\ 1983, ApJ
264, 152
\re
Binney, J., Merrifield, M.\ 1998, Galactic Astronomy, Princeton Series in
Astrophysics (Princeton, New Jersey)
\re
Bryant, P.M., Scoville, N.Z.\ 1999, ApJ 117, 2632
\re
Colbert, E.J.M., Wilson, A.S., Bland-Hawthorn, J.\ 1994, ApJ 436, 89
\re
Condon, J.J., Condon, M.A., Gisler, G., Puschell, J.J.\ 1982, ApJ 252, 102
\re
Cowie, L.L., McKee, C.F., Ostriker, J.P.\ 1981, ApJ 247, 908
\re
DePoy, D.K., Becklin, E.E., Wynn-Williams, C.G.\ 1986, ApJ 307, 116
\re
Doyon, R., Wells, M., Wright, G.S., Joseph, R.D., Nadeau, D., James, P.A.\ 1994,
ApJL 437, 23
\re
Draine, B.T., Woods, D.T.\ 1990, ApJ 363, 464
\re
Eales, S.A., Becklin, E.E., Hodapp, K.-W., Simons, D.A., Wynn-Williams,
C.G.\ 1990, ApJ 365, 478
\re
Elston, R., Maloney, P.\ 1990, ApJ 357, 91
\re
Fried, J.W., Schulz, H.\ 1983, A\&A 118, 166
\re
Garwood, R.W., Helou, G., Dickey, J.M.\ 1987, ApJ 322, 88
\re
Goldader, J.D., Joseph, R.D., Doyon, R., Sanders, D.B.\ 1995, ApJ 444, 97
\re
Heckman, T.M., Balick, B., van Breugel, W.J.M., Miley, G.K.\ 1983, AJ 88, 583
\re
Heckman, T.M., Armus, L., Miley, G.K.\ 1987, AJ 92, 276
\re
Heckman, T.M., Armus, L., Miley, G.K.\ 1990, ApJS 74, 833
\re
Heckman, T.M., Lehnert, M.D., Armus, L.\ 1993, in The Environment and
Evolution of Galaxies, ed J.M.\ Shull, H.A.\ Thronson Jr.\ (Kluwer, Netherlands)
p455
\re
Heckman, T.M., Dahlem, M., Eales, S.A., Fabbiano, G., Weaver, K.\ 1996, ApJ 457,
616
\re
Herbst, T.M., Graham, J.R., Beckwith, S., Tsutsui, K., Soifer, B.T.,
Matthews, K.\ 1990, AJ 99, 1773
\re
Iwasawa, K., Comastri, A.\ 1998, MNRAS 297, 1219
\re
Joseph, R.D., Wright, G.S.\ 1985, MNRAS 214, 87
\re
Kaifu, N.\ 1998, SPIE 3352, 14
\re
Keel, W.C.\ 1990, AJ 100, 356
\re
Klaas, U., Hass, M., Heinrichsen, I., Schulz, B.\ 1997, A\&A 325, L21
\re
Kronberg, D P., Biermann, P., Schwab, F.R.\ 1985, ApJ 291, 693
\re
Lester, D.F., Gaffney, N.I.\ 1994, ApJL 431, 13
\re
Lester, D.F., Harvey, P.M., Carr, J.\ 1988, ApJ 329, 641
\re
Moorwood, A.F.M., Oliva, E.\ 1990, A\&A 239, 78
\re
Motohara, K., Maihara, T., Iwamuro, F., Oya, S., Imanishi, M., Terada, H.,
Goto, M., Iwai, J.\ et al.\ 1998, SPIE 3354, 659
\re
Mouri, H.\ 1994, ApJ 427, 777
\re
Mouri, H., Taniguchi, Y.\ 1992, ApJ 386, 68
\re
Mouri, H., Taniguchi, Y.\ 1995, ApJ 449, 134
\re
Mouri, H., Taniguchi, Y., Kawara, K., Nishida, M.\ 1989, ApJL 346, 73
\re
Nakai, N., Hayashi, M., Handa, T., Sofue, Y., Hasegawa, T., Sasaki M.\ 1987,
PASJ 39, 685
\re
Ohyama, Y., Taniguchi, Y.\ 1998, ApJL 499, 934
\re
Osterbrock, D.E.\ 1989, Astrophysics of Gaseous Nebulae and Active Galactic
Nuclei (Mill Valley, California)
\re
Phillips, A.C.\ 1993, AJ 105, 486
\re
Rieke, G.H., Lebofsky, M.J., Thompson, R.I., Low, F.J., Tokunaga, A.T.\ 1980,
ApJ 238, 24
\re
Rieke, G.H., Cutri, R.M., Black, J.H., Kailey, W.F., McAlary, C.W.,
Lebofsky, M.J., Elston, R.\ 1985, ApJ 290, 116
\re
Sanders, D.B., Mirabel, I.F.\ 1985, ApJL 298, 31
\re
Sanders, D.B., Mirabel, I.F.\ 1996, ARA\&A 34, 749
\re
Smith, C.H., Aitken, D.K., Roche, P.F.\ 1989, MNRAS 241, 425
\re
Solomon, P.M., Downes, D., Radford, S.J.E., Barrett, J.W.\ 1997, ApJ 478, 144
\re
Suchkov, A.A., Balsara, D.S., Heckman, T.M., Leitherer, C.\ 1994, ApJ 430, 511
\re
Sugai, H., Malkan, M.A., Ward, M.J., Davies, R.I., McLean, I.S.\ 1997, ApJ
481, 186
\re
Tanaka, M., Hasegawa, T., Gatley, I.\ 1991, ApJ 374, 516
\re
Tanaka, M., Hasegawa, T., Hayashi, S.S., Brand, P.W.J.L., Gatley, I.\ 1989, ApJ
336, 207
\re
Tacconi, L.J., Genzel, R., Tecza, M., Gallimore, J.F., Downes, D.,
Scoville, N.Z.\ 1999, astro-ph/9905031
\re
Thronson, H.A.\ Jr, Majewski, S., Descartes, L., Hereld, M.\ 1990, ApJ 364, 456
\re
Tomisaka, K., Ikeuchi, S.\ 1988, ApJ 330, 695
\re
van der Werf, P.P., Genzel, R., Krabbe, A., Blietz, M., Lutz, D., Drapatz, B.,
Ward, M.J., Forbes, D.A.\ 1993, ApJ 405, 522
\re
Wang, Z., Scoville, N.Z., Sanders, D.B.\ 1991, ApJ 368, 112
\re
Wright, G.S., Joseph, R.D., Meikle, W.P.S.\ 1984, Nature 309, 430
\re
Yoshida, M., Taniguchi, Y., Murayama, T.\ 1999, AJ 117, 1158

%------------------------------------------------------------------------------
% Table 1
%------------------------------------------------------------------------------

% table 1
\begin{table*}[t]
\begin{center}
Table~1. \hspace{4pt}Previous measurements of systemic velocity of NGC 6240.\\
\end{center}
\vspace{6pt}
\begin{tabular*}{\textwidth}{@{\hspace{\tabcolsep}
\extracolsep{\fill}}p{6pc}ccl}
\hline\hline\\[-6pt]
$V_\solar$ (km s$^{-1}$)& Waveband & Comments & Reference \\[4pt]\hline\\[-6pt]\\
$7287$ & HI 21 cm & Absorption, Single dish & Garwood, Helou, \& Dickey (1987) \\
$7350\pm 65$ & HI 21 cm & Absorption, Single dish & Heckman et al. (1983) \\
$7270$ & HI 21 cm & Absorption, Single dish & Baan et al. (1985) \\
$7253$ & OH 1667, 1667 MHz & Absorption, Single dish & Baan et al. (1985) \\
$7285$ & $^{12}$CO($J=1-0$) & Single dish observation & Sanders \& Mirabel (1985) \\
$7298$ & $^{12}$CO($J=1-0$) & Single dish & Solomon et al. (1997) \\
$7335\pm 13$ & $^{12}$CO($J=1-0$) & Interferometer & Bryant \& Scoville (1999) \\
$7313$ & $^{12}$CO($J=1-0$) & Interferometer & Wang et al. (1991) \\
$7297$ & $1-0~S$(1) & Estimated from van der Werf et al.'s figure 3 & van der Werf et al. (1993) \\
$7275\pm 50$ & Stellar CO absorption & K band & Lester \& Gaffney (1994) \\
$7473\pm 30$ & Stellar CO absorption & K band & Doyon et al. (1994) \\
\hline
\end{tabular*}
\end{table*}

%------------------------------------------------------------------------------
% Figures
%------------------------------------------------------------------------------

\clearpage
\centerline{Figure Captions}
\bigskip
%
% Fig. 1
\begin{fv}{1}
{7cm}{
The K band image of NGC 6240.
This image was obtained through a wide $2''$ slit with its
position angle of $19^\circ$ and is is averaged over 2 pixels
($0.\hspace{-2pt}''23$).
No correction for the dark subtraction and the flatfielding were
applied since this image is taken just for the source acquisition.
The coordinates are shown relative to the S nucleus in units of
arcsecond.
The spatial resolution is indicated with a circle at the lower left
corner.
Two white vertical lines indicate the slit position used for the
spectroscopy ($0.\hspace{-2pt}''5$ width).
The peak position of $^{12}$CO($J=2-1$) (Tacconi et al.\ 1999) is
marked with a cross sign.
This position is corrected for $0.\hspace{-2pt}''15$ offset between the radio
and the K band nuclei (see the main text for the detail).
}
\end{fv}

% Fig. 2
\begin{fv}{2}
{7cm}{
Spectra extracted over 2 pixels ($0.\hspace{-2pt}''23$) at various positions
along the slit with the line identifications.
Relative offset from the S nucleus are indicated at each spectrum in
units of arcsecond as well as the positions of the N and the S nuclei.
All spectra are normalized with the peak flux of the $1-0~S$(1)
emission.
Zero flux levels are shifted for clarity and are indicated at the left
ordinate for each spectrum.
Some data points are removed since they are affected by bad pixels.
}
\end{fv}

% Fig. 3
\begin{fv}{3}
{7cm}{
$1-0~S$(1) flux and velocity field along the slit.
Positions of the N and the S nuclei are indicated at the top of the
panels.
1$\sigma$ errors are shown for the all plots.
($top$)
$1-0~S$(1) flux, continuum flux at $2.2\mu$m, and equivalent width of
$1-0~S$(1) (right ordinate) are shown with squares, circles, and
crosses, respectively.
($middle$)
Heliocentric velocity of the $1-0~S$(1) emission measured with a
single Gaussian function.
($bottom$)
Line width (FWHM) of the $1-0~S$(1) emission measured with a single
Gaussian function.
}
\end{fv}

% Fig. 4
\begin{fv}{4}
{7cm}{
Velocity field of the $1-0~S$(1) emission.
Middle image shows a spectroscopic image of $1-0~S$(1).
In order to show the velocity structure clearly, the image is
processed in the following ways.
First, the continuum flux is removed by interpolating the adjacent
continuum.
Then, the peak of the resultant emission-line spectroscopic image is
normalized to be unity at every row.
Note that the contour levels shown in this image represent neither the
relative flux level nor the significance of the flux as a result of
the peak normalization.
You can find the significance of the contour levels with the plots of
the continuum and $1-0~S$(1) flux in the left panel.
The emission-line profile of the $1-0~S$(1) emission at three
representative positions (at three horizontal bars in the image marked
as a, b, and c) are shown in the right panels.
Typical errors of the profiles are indicated with vertical bars in
each panel.
}
\end{fv}

% Fig. 5
\begin{fv}{5}
{7cm}{
The observed line profile of the $1-0~S$(1) emission at $7''$
south of the S nucleus is compared with the model line profile.
The continuum emission is removed for the observed profile.
The blueshifted and redshifted model components are shown with a
dotted and a short-dashed lines, respectively, and the sum of the two
components is plotted with a long-dashed line.
See the main text for the model description.
}
\end{fv}

% Fig. 6
\begin{fv}{6}
{7cm}{
The line ratio vs. line ratio diagram.
$1-0~S$(2)/$1-0~S$(0) ratio is plotted against $2-1~S$(1)/$1-0~S$(1)
ratio with 1$\sigma$ errors.
The locus shows the theoretical line ratios of the thermal excitation
with the temperature $< 3000$ K (marked along the curve).
Two open squares are observed data points for supernova remnants IC
443 and RCW 103 taken from Mouri (1994).
}
\end{fv}

% Fig. 7
\begin{fv}{7}
{7cm}{
The population diagrams.
Populations relative to ($J, v$)=(3, 1) state are plotted for (2, 1),
(4, 1), (3, 2), and (5, 2) (corresponding to $1-0~S$(0), $1-0~S$(2),
$2-1~S$(1), and $2-1~S$(3) emission lines, respectively) at every
$0.\hspace{-2pt}''23$ with 1$\sigma$ errors.
Each diagram is shifted for clarity and the diagrams at the most
southern and northern regions are plotted at the upper left and the
lower right corners of the plots.
Positions of the N and the S nuclei are indicated.
Only lower limits of the populations of (2, 1) state are measured
around the S nucleus because of the weak $1-0~S$(0) emission on the
strong continuum emission with absorption lines.
}
\end{fv}

% Fig. 8
\begin{fv}{8}
{7cm}{
Spatial variation of the $v=2-1$ vibration temperature ($T_{\rm vib}$)
and $v=1$ rotation temperature ($T_{\rm rot}$).
1$\sigma$ errors are shown for the all plots.
Positions of the N and the S nuclei are indicated.
($top$)
The ratio of $T_{\rm vib}$/$T_{\rm rot}$ is plotted along the slit.
The two dotted horizontal lines at 1 and 1.25 indicate the range
expected for the thermal excitation (see the main text for the
detail).
($middle$)
The $2-1~S$(1)/$1-0~S$(1) flux ratio (left ordinate) and $T_{\rm vib}$
(right ordinate) are shown along the slit.
($bottom$)
The $1-0~S$(2)/$1-0~S$(0) flux ratio (left ordinate) and $T_{\rm rot}$
(right ordinate) are shown along the slit.
}
\end{fv}

% Fig. 9
\begin{fv}{9}
{7cm}{
Our proposed scenario of NGC 6240 is schematically shown.
($a$)
Schematic picture at the early stage of the merging with nuclear
starburst and molecular gas transfer.
See the main text for the detail.
($b$)
Schematic picture at the late stage of the merging with the superwind
and the intense H$_2$ emission.
See the main text for the detail.
}
\end{fv}

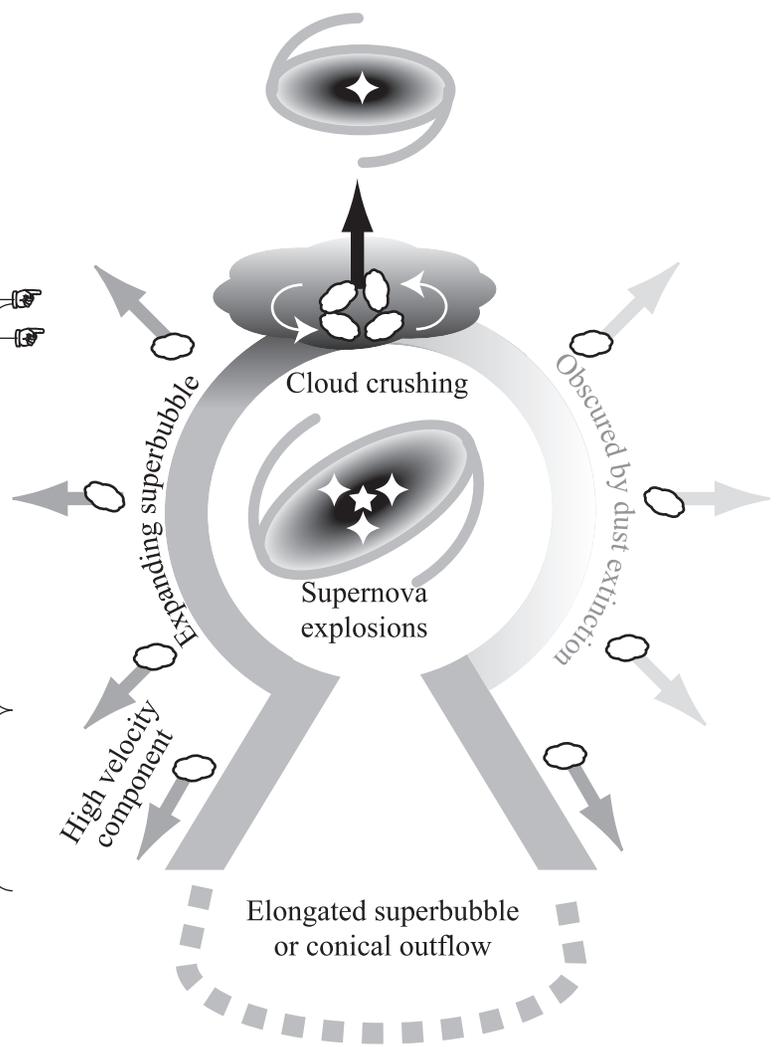
\end{document}